\documentclass[aps,prd,preprintnumbers,superscriptaddress,nofootinbib,notitlepage,twocolumn]{revtex4-2}
\usepackage[pdftex]{graphicx}
\usepackage{bm,latexsym,amsmath,amssymb,amsfonts,mathrsfs}
\usepackage{color}
\allowdisplaybreaks[1]
\usepackage[pdftex,colorlinks=true,linkcolor=blue,citecolor=cyan]{hyperref}
\newcommand*{\D}{{\rm d}}
\newcommand*{\mpl}{M_{\rm Pl}}

\begin{document}
\title{Stars disformally coupled to a shift-symmetric scalar field}
\author{Tact~Ikeda}
\email[Email: ]{tact@rikkyo.ac.jp}
\affiliation{Department of Physics, Rikkyo University, Toshima, Tokyo 171-8501, Japan
}
\author{Aya~Iyonaga}
\email[Email: ]{iyonaga@rikkyo.ac.jp}
\affiliation{Department of Physics, Rikkyo University, Toshima, Tokyo 171-8501, Japan
}
\author{Tsutomu~Kobayashi}
\email[Email: ]{tsutomu@rikkyo.ac.jp}
\affiliation{Department of Physics, Rikkyo University, Toshima, Tokyo 171-8501, Japan
}
\begin{abstract}
We investigate static and spherically symmetric stars disformally
coupled to a scalar field. The scalar field is assumed to be shift symmetric,
and hence the conformal and disformal factors of the metric coupled to matter
are dependent only on the kinetic term of the scalar field.
Assuming that the scalar field is linearly dependent on time,
we consider a general shift-symmetric scalar-tensor theory and
a general form of the matter energy-momentum tensor
that allows for the anisotropic pressure and the heat flux in the radial direction.
This is a natural starting point in light of how
the gravitational field equations and
the energy-momentum tensor transform under a disformal transformation.
By inspecting the structure of the hydrostatic equilibrium equation
in the presence of the derivative-dependent conformal and disformal factors,
we show that the energy density and tangential pressure must vanish
at the surface of a star. This fact is used to prove the disformal invariance of
the surface of a star, which was previously subtle and unclear.
We then focus on the shift-symmetric k-essence disformally coupled to matter,
and study the interior and exterior metric functions and
scalar-field profile in more detail.
It is found that there are two branches of the solution
depending on the velocity of the scalar field.
The disformally-related metric functions
of the exterior spacetime are also discussed.
\end{abstract}
\preprint{RUP-21-13}
\maketitle
\section{Introduction}

The accelerated expansion of the present Universe
is implied by the observations of Type Ia supernovae~\cite{Riess:1998cb,Perlmutter:1998np}.
The source of this accelerated expansion, however, remains unknown
and is called generically dark energy.
The intriguing question here is
whether the dark energy component is in fact just a cosmological constant or
some dynamical fields (see, e.g., Ref.~\cite{Copeland:2006wr} for a review).
In many models, dark energy is assumed to be described by
a scalar field, and the simplest examples include quintessence~\cite{Fujii:1982ms,Ford:1987de,Wetterich:1987fm,Chiba:1997ej,Caldwell:1997ii}
and k-essence~\cite{ArmendarizPicon:2000dh,Chiba:1999ka}.
If the accelerated expansion were caused by such a dynamical scalar field,
it might be coupled to
(a part of) matter fields, giving rise to signatures of the fifth force.
For example, matter may be coupled to a dark energy
scalar field $\phi$ through the metric $\widetilde g_{\mu\nu}$
which is conformally related to the Einstein-frame metric $g_{\mu\nu}$
as $\widetilde g_{\mu\nu}=\Omega(\phi)g_{\mu\nu}$,
where the conformal factor $\Omega$ is a function of $\phi$.
A more general way of non-minimal coupling is obtained from
a disformally related metric~\cite{Bekenstein:1992pj},
\begin{align}
	\widetilde g_{\mu\nu}=\Omega(\phi,X)g_{\mu\nu}+\Gamma(\phi,X)\phi_\mu\phi_\nu,\label{intro:disf}
\end{align}
where $\phi_\mu:=\partial_\mu\phi$ and the conformal and disformal factors $\Omega$ and $\Gamma$
may depend on
$X:=-g^{\mu\nu}\phi_\mu\phi_\nu/2$
as well as $\phi$.
This coupling stems from the most general metric transformation
that includes up to first-order derivatives of a scalar field.

Another widely studied approach to
account for the accelerated expansion of the Universe
is modifying gravity on cosmological scales.
Many modified gravity theories can be described
by scalar-tensor theories at least effectively in a certain limit.
A scalar-field theory for dark energy in the presence of
disformally coupled matter can be recast into a scalar-tensor theory
minimally coupled to matter through a disformal transformation~\eqref{intro:disf}.
Conversely, one can perform a transformation~\eqref{intro:disf}
to rewrite a given scalar-tensor theory to
a scalar-field model of dark energy or another scalar-tensor theory.
It is just a matter of the frames as long as the transformation is invertible.

Among various theories, scalar-tensor theories
in the Horndeski family~\cite{Horndeski:1974wa,Deffayet:2011gz,Kobayashi:2011nu}
have the healthy property that the field equations are of second order and hence
the Ostrogradsky instability can be avoided~\cite{Ostrogradsky:1850fid,Woodard:2006nt,Woodard:2015zca}.
Recently, more general healthy scalar-tensor theories
have been developed named degenerate higher-order scalar-tensor (DHOST)
theories~\cite{Zumalacarregui:2013pma,Langlois:2015cwa,Crisostomi:2016czh,Klein:2016aiq} (see Refs.~\cite{Langlois:2018dxi,Kobayashi:2019hrl} for a review).
The field equations in DHOST theories are of higher order,
but there appear no dangerous Ostrogradsky
modes thanks to the degenerate nature
of the system.
The disformal transformation of the metric
plays an important role in DHOST theories:
each subclass of DHOST theories is stable under
transformations of the form~\eqref{intro:disf},\footnote{The
Horndeski class is stable
under $X$-independent disformal transformations~\cite{Bettoni:2013diz}.}
and, in particular, the Lagrangian in
the physically interesting subclass called the ``class Ia''
can always be transformed to the one in the Horndeski family
via a disformal
transformation~\cite{Crisostomi:2016czh,Achour:2016rkg}.\footnote{See Ref.~\cite{Langlois:2020xbc} for further update on transformation properties from
the viewpoint of three-dimensional geometric quantities.}
Therefore, a DHOST theory with minimally coupled matter
can be mapped to a Horndeski theory with disformally coupled matter.
In vacuum, disformal transformations can be used
as a solution-generating method for finding
new solutions in DHOST theories~\cite{BenAchour:2019fdf,Anson:2020trg,BenAchour:2020fgy,Minamitsuji:2020jvf,Faraoni:2021gdl,Achour:2021pla}.

Given the importance of the disformal metric in dark energy
and modified gravity models, it is natural to explore
cosmological and astrophysical implications of disformal coupling to matter.
The disformal coupling has been studied in the context of cosmology
in Refs.~\cite{Kaloper:2003yf,Koivisto:2008ak,Zumalacarregui:2010wj,Zumalacarregui:2012us,Koivisto:2012za,Koivisto:2013fta,Sakstein:2014aca,Sakstein:2014isa,Sakstein:2015jca,vandeBruck:2015ida, Hagala:2015paa,vandeBruck:2015tna}.
The invariance of cosmological observables under disformal transformations
has been discussed in Refs.~\cite{Minamitsuji:2014waa,Tsujikawa:2014uza,Domenech:2015hka,Motohashi:2015pra,Chiba:2020mte}.
Disformal couplings to baryons and photons have been considered
in Refs.~\cite{Kaloper:2003yf,Brax:2014vva,Brax:2014vla,Brax:2014zba,Brax:2015hma,Lamm:2015gka},
while couplings to the dark sector have been studied in Refs.~\cite{Neveu:2014vua,Chibana:2019jrf}.
Other applications can be found in
Refs.~\cite{Ip:2015qsa,Koivisto:2015mwa,vandeBruck:2016vlw,Andreou:2019ikc,Silva:2019rle,Ramazanoglu:2019jrr,Ramazanoglu:2019jfy,Erices:2021uyu,Brax:2021qqo}.
A multi-field extension of disformal transformations has been explored
in Ref.~\cite{Watanabe:2015uqa}.

In this paper, we study static and spherically symmetric
objects in the presence of matter disformally coupled to a scalar field.
Relativistic stars in the presence of disformal coupling
have been studied in Ref.~\cite{Minamitsuji:2016hkk},
where the two coupling functions depend only on $\phi$
and not on $X$. Very recently, properties of neutron stars
in the presence of $X$-dependent conformal coupling
have been discussed~\cite{Boumaza:2021hzr}.
Disformal transformations of various quantities concerning
relativistic stars have been investigated in Ref.~\cite{Minamitsuji:2021rtw}.
Our basic motivation is the same as these works.
In contrast to the above works, however, we assume
the shift symmetry, $\phi\to\phi+c$~($c=\,$const.), so that the conformal and disformal
factors as well as the gravitational sector of the theory
depend on the scalar field only through its derivatives.
Under this assumption we consider the linearly time-dependent
ansatz for the scalar field. Our setup is similar to,
but different from ``derivative chameleons''~\cite{Noller:2012sv}
in various aspects.
In the latter half of the paper, we focus on
a k-essence field disformally coupled to matter.
By making a disformal transformation, our system is mapped
to a certain DHOST theory in the presence of minimally coupled matter.
A distinct point is that the Vainshtein screening mechanism
is not expected to operate in that particular DHOST theory,
as opposed to the case of generic DHOST
theories~\cite{Kobayashi:2014ida,Saito:2015fza,Babichev:2016jom,Crisostomi:2017lbg,Langlois:2017dyl,Dima:2017pwp,Chagoya:2018lmv,Kobayashi:2018xvr,Hirano:2019scf,Crisostomi:2019yfo,Anson:2020fum}.

This paper is organized as follows.
In the next section, we overview the gravitational field equations
and fluid equations for a static and spherically symmetric object
disformally coupled to a shift-symmetric scalar field.
We investigate the hydrostatic equilibrium equation in more detail
and show that, in contrast to the case of minimal coupling,
the density and tangential pressure must be continuous across
the surface of a star. This is the key to prove the disformal
invariance of the stellar surface.
In Sec.~\ref{sec:caseofk}, we focus on the shift-symmetric k-essence
theory disformally coupled to matter and study
the interior and exterior solutions of a spherical object.
Finally, we draw our conclusions in Sec.~\ref{sec:concl}.

\section{Spherically symmetric stars disformally coupled to a scalar field}\label{sec:stars}

In this section, we first derive general field equations
in the case where a scalar field is disformally coupled to matter,
and then provide basic equations for the study of spherically symmetric stars
in the presence of a disformally coupled scalar field with shift symmetry.

\subsection{Field equations}

We start with a general class of scalar-tensor theories
described by the action
\begin{align}
    S&=S_{\rm grav}[g_{\mu\nu},\phi] +
    S_{\rm m}[\widetilde g_{\mu\nu},\psi_{\rm m}]
    \notag \\
    &=\int \D^4x\sqrt{-g}{\cal L}_{\rm grav}(g_{\mu\nu},\phi)
    +\int \D^4x\sqrt{-\widetilde g}{\cal L}_{\rm m}(\widetilde g_{\mu\nu},\psi_{\rm m}).
    \label{eq:action0}
\end{align}
where the matter field $\psi_{\rm m}$ is coupled to
the disformally-related metric~\eqref{intro:disf}.

Varying the action~\eqref{eq:action0} with respect to the metric $g_{\mu\nu}$,
we obtain the gravitational field equations,
\begin{align}
    \mathcal{E}^{\mu \nu}:=\mathcal{G}^{\mu \nu}-T^{\mu \nu}=0,
\end{align}
where 
\begin{align}
    \mathcal{G}^{\mu \nu}=-\frac{2}{\sqrt{-g}} \frac{\delta S_{\rm grav}}{\delta g_{\mu \nu}},
    \quad
    T^{\mu \nu}=\frac{2}{\sqrt{-g}} \frac{\delta S_{\rm m}}{\delta g_{\mu \nu}}.
\end{align}
Varying the action~\eqref{eq:action0} with respect to the
scalar field $\phi$, we obtain the scalar-field equation of motion,
\begin{align}
    \mathcal{E}_\phi :=\mathcal{F}_\phi-T_\phi =0,
\end{align}
where
\begin{align}
    \mathcal{F}_{\phi}=\frac{1}{\sqrt{-g}} \frac{\delta S_{\rm grav}}{\delta \phi},
    \quad
    T_\phi =-\frac{1}{\sqrt{-g}}\frac{\delta S_{\rm m}}{\delta \phi}.
\end{align}

We assume that $S_{\rm grav}$ and $S_{\rm m}$ are separately
invariant under an infinitesimal coordinate transformation,
$x^\mu\to x^\mu+\xi^\mu$.
It then follows the Bianchi identities,
\begin{align}
    \nabla_\nu {\cal E}^{\mu\nu}+{\cal E}_\phi \phi^\mu &=0,\label{bia}
    \\
    \nabla_\nu T^{\mu\nu}+T_\phi \phi^\mu &=0.\label{cons}
\end{align}
Equivalently, one may use
$\nabla_\nu {\cal G}^{\mu\nu}+{\cal F}_\phi \phi^\mu =0$
instead of Eq.~\eqref{bia}.

Using the fact that $S_{\rm m}$ depends on $g_{\mu\nu}$ and $\phi$
through the disformally-related metric~\eqref{intro:disf},
we write $T_\phi$ in a more explicit form:
\begin{align}
    T_\phi = \nabla_\mu W^\mu - Z,\label{eq:Tphi=WZ}
\end{align}
where
\begin{align}
    W^\mu &=w_1 T^{\mu\nu}\phi_\nu + w_2 T\phi^\mu 
    + w_3 T^{\alpha\beta}\phi_\alpha\phi_\beta\phi^\mu, \label{eq:wmTTT}
    \\
    Z&=z_1 T+z_2T^{\alpha\beta}\phi_\alpha\phi_\beta,
\end{align}
and $T=T_\mu^{\mu}$,
with $w_i$ and $z_j$ being functions of $\phi$ and $X$ defined as
\begin{align}
    w_1&=\frac{\Gamma}{\Omega},\label{eq:defw1}
    \\
    w_2&=-\frac{1}{2}
    \frac{\Omega_X}{\Omega}\frac{\Omega-2X\Gamma}{\Omega-X\Omega_X+2X^2\Gamma_X},
    \\
    w_3&=-\frac{1}{2}
    \frac{\Gamma_X}{\Omega}\frac{\Omega-2X\Gamma}{\Omega-X\Omega_X+2X^2\Gamma_X},
    \\
    z_1&=\frac{1}{2}\left(\frac{\Omega_\phi}{\Omega}+\frac{X\Omega_X}{\Omega}
    \frac{\Omega_\phi-2X\Gamma_\phi}{\Omega-X\Omega_X+2X^2\Gamma_X}\right),
    \\
    z_2&=\frac{1}{2}\left(\frac{\Gamma_\phi}{\Omega}+\frac{X\Gamma_X}{\Omega}
    \frac{\Omega_\phi-2X\Gamma_\phi}{\Omega-X\Omega_X+2X^2\Gamma_X}
    \right).\label{eq:defz2}
\end{align}
A detailed derivation is presented in Appendix~\ref{app:basic-eqs}
(see also Ref.~\cite{Chibana:2019jrf}).
Note that if the coupling functions $\Omega$ and $\Gamma$
preserve the shift symmetry, $\phi\to\phi+c$, then one has $Z=0$.

\subsection{Basic equations for spherically symmetric stars}

Hereafter, we focus on shift-symmetric theories,
$\phi\to\phi+c$, and require the same symmetry for the coupling functions.

Let us consider a static and spherically symmetric metric,
\begin{align}
    \D s^2=-e^{\nu (r)}\D t^2+e^{\lambda (r)}\D r^2
    +r^2\left(\D \theta^2+\sin^2 \theta \D\varphi^2\right).
\end{align}
Even though the metric is assumed to be static,
the shift symmetry admits a linearly time-dependent configuration of the scalar field,
\begin{align}
	\phi=\mu t+\psi(r),
\end{align}
where $\mu$ ($\neq0$) is a constant.

We assume the following general form of the energy-momentum tensor,
\begin{align}
    T^\mu_\nu=\left(
\begin{array}{cccc}
 -\rho(r) & e^{-\nu}\chi(r) & 0 & 0 \\
 -e^{-\lambda}\chi(r) & P_r(r) & 0 & 0 \\
 0 & 0 & P_\perp(r) & 0 \\
 0 & 0 & 0 & P_\perp(r) \\
\end{array}
\right),\label{energy-momentum}
\end{align}
where we allow for the off-diagonal component, $\chi(r):=-T_{tr}$,
i.e., the heat flux in the radial direction,
and anisotropic pressure, $P_r\neq P_\perp$.
Below we will see that in general this off-diagonal component is
necessary if $\mu\neq 0$.

Under the above ansatz, the radial and temporal components of
the Bianchi identities~\eqref{bia} read, respectively,
\begin{align}
    \frac{e^{-\nu/2}}{r^2}\left(r^2e^{\nu/2}{\cal E}_r^r\right)'
    -\frac{\nu'}{2}{\cal E}_t^t 
    -\frac{2}{r}{\cal E}_\perp+{\cal E}_\phi\psi'&=0,\label{eq:bia1}
    \\
    \frac{e^{-(\nu+\lambda)/2}}{r^2}\left[r^2e^{(\nu+\lambda)/2}{\cal E}_t^r\right]'
    +{\cal E}_\phi\mu &=0,\label{eq:bia2}
\end{align}
where a prime denotes differentiation with respect to $r$
and ${\cal E}_\perp$ stands for the angular components,
${\cal E}_\perp:={\cal E}_\theta^\theta={\cal E}_\varphi^\varphi$.
These identities show that one may take 
\begin{align}
    &\mathcal{E}^t_t={\cal G}_t^t+\rho=0,
    \\
    &\mathcal{E}^r_r={\cal G}_r^r-P_r=0,
    \\
    &\mathcal{E}_{tr}={\cal G}_{tr}+\chi=0,
\end{align}
as independent equations, while ${\cal E}_\perp=0$ and ${\cal E}_\phi=0$
follow automatically from the Bianchi identities.

We now look into Eq.~\eqref{cons}. Similarly to Eqs.~\eqref{eq:bia1}
and~\eqref{eq:bia2}, we obtain
\begin{align}
    P_r'+\frac{\nu'}{2}(\rho+P_r)+\frac{2}{r}(P_r-P_\perp)+T_\phi\psi'&=0,
    \label{eq:hydrostatic}
    \\
    \frac{e^{-(\nu+\lambda)/2}}{r^2}\left[r^2e^{(\nu+\lambda)/2}T_t^r\right]'
    +T_\phi\mu &=0,\label{eq:dchiWrdr}
\end{align}
where, by noting that $Z=0$ due to the shift symmetry,
\begin{align}
    T_\phi = \frac{e^{-(\nu+\lambda)/2}}{r^2}\left[r^2e^{(\nu+\lambda)/2}W^r\right]'.
\end{align}
Explicitly, one has
\begin{align}
    W^r&=
    e^{-\lambda}\psi'
    \bigl\{
    w_1P_r-w_2(\rho-P_r-2P_\perp)
    \notag \\ &\quad 
    +w_3\left[
    \mu^2e^{-\nu}\rho+e^{-\lambda}(\psi')^2P_r
    \right]
    \bigr\}
    \notag \\ &\quad 
    +\mu e^{-\nu-\lambda}\left[w_1+2w_3e^{-\lambda}(\psi')^2\right]\chi.
\end{align}
One can immediately integrate Eq.~\eqref{eq:dchiWrdr} to get
\begin{align}
    e^{-\lambda} \chi-\mu W^r=\frac{{\cal C}e^{-(\nu+\lambda)/2}}{r^2},
    \label{eq:intchiW}
\end{align}
where ${\cal C}$ is an integration constant.
To fix ${\cal C}$, let us consider the behavior of the left-hand side
at the center. We may assume that
$\nu$, $\lambda$, $\psi'$, $\rho$, $P_r$, $P_\perp$, and $\chi$
do not diverge as $r\to 0$. Then, $W^r$ does not diverge
at the center. This observation leads us to set ${\cal C}=0$.
Having shown that, we can solve Eq.~\eqref{eq:intchiW}
for $\chi$ to obtain
\begin{align}
    \chi = \mu \psi'h(\nu,\lambda,\psi',\rho,P_r,P_\perp),\label{eq:chih01}
\end{align}
where
\begin{align}
    h&:=\frac{
    w_1P_r-w_2(\rho-P_r-2P_\perp)
    }{1-\mu^2e^{-\nu}[w_1 +2w_3e^{-\lambda}(\psi')^2]}
    \notag \\ &\quad 
    +\frac{
    w_3\left[
    \mu^2e^{-\nu}\rho+e^{-\lambda}(\psi')^2P_r
    \right]
    }{1-\mu^2e^{-\nu}[w_1 +2w_3e^{-\lambda}(\psi')^2]}
    .
\end{align}
In general, there is no reason to impose $h=0$ and
hence we have $\chi\neq 0$.

\subsection{The surface of a star}

We now consider more specifically
the case where ${\cal L}_{\rm grav}$ is given by
that of the shift-symmetric Horndeski theory,
\begin{align}
{\cal L}_{\rm grav}&=G_2(X)-G_3(X)\Box\phi
+G_4(X)R
\notag \\ & \quad
+G_{4X}\left[(\Box\phi)^2-\nabla_\mu\nabla_\nu\phi\nabla^\mu\nabla^\nu\phi
\right].
\end{align}
We do not consider the ``$G_5$'' terms for simplicity;
they do not lead to no essential difference in the following results.

The goal of this subsection is to discuss the matching conditions
at the surface of a star, $r=r_s$, defined by  $P_r(r_s)=0$.
For this purpose only the structure of
the highest derivative terms in ${\cal G}_{\mu\nu}$
is important.
We have
\begin{align}
{\cal G}_t^t&=g_1+g_2\lambda'+g_3\psi''=-\rho,\label{horn:tt}
\\
{\cal G}_r^r&=g_4-g_2\nu'=P_r,\label{horn:rr}
\\
{\cal G}_{tr}&=\frac{\mu e^\lambda}{2}\left(g_5+g_3\nu'\right)=-\chi,\label{horn:tr}
\end{align}
where $g_a=g_a(\nu,\lambda,\psi')$ ($a=1,2,3,4,5$).
Using Eqs.~\eqref{horn:rr} and~\eqref{horn:tr} one can eliminate $\nu'$
to get the equation of the form
\begin{align}
    \lambda=F_\lambda (\nu,\psi',P_r,\chi).\label{horn:lambdasol}
\end{align}
Equation~\eqref{eq:hydrostatic} is independent of
the specific form of the gravitational part of the Lagrangian.

We impose that the induced metric
is continuous across the surface of a star,
and hence $\nu$ is continuous.
If the matter is minimally coupled to gravity,
$\rho$ and $P_\perp$ are allowed to be discontinuous at $r=r_s$,
and then it follows from Eqs.~\eqref{horn:tt} and~\eqref{horn:lambdasol} that
$\lambda'$ and $\psi''$ may be discontinuous. 
The situation becomes more subtle if the matter is non-minimally
coupled to gravity through $X$-dependent $\Omega$ and $\Gamma$.
Now, we have
\begin{align}
    T_\phi\supset \frac{
    e^{-\lambda}\psi'\left[
    -(w_2-w_3\mu^2e^{-\nu})\rho'+2w_2P_\perp'
    \right]
    }{1-\mu^2e^{-\nu}[w_1 +2w_3e^{-\lambda}(\psi')^2]},
\end{align}
which yields the delta functions $\sim\delta(r-r_s)$.
However, we see that no other terms in Eq.~\eqref{eq:hydrostatic}
contain the delta functions canceling $T_\phi$.
It is therefore required in the present case that
$\rho$ and $P_\perp$ are continuous across the surface, i.e.,
$\lim_{r\to r_s-0}\rho(r)=0$ and $\lim_{r\to r_s-0}P_\perp (r)=0$.
Note that the first derivatives of $\rho$, $P_r$, and $P_\perp$
may be discontinuous at $r=r_s$, and so are $\lambda'$ and $\psi''$,
provided that $g_2\lambda' + g_3\psi''$ is continuous.
Note also that there is an exceptional case where
$\psi'(r_s)=0$, to which the above argument does not apply.

The requirement that $\rho(r_s)=P_\perp(r_s)=0$
in the presence of the $X$-dependent coupling functions
has a direct implication on the disformal invariance of
the notion of the surface of a star.
As discussed in Appendix~\ref{appB-dtemt},
the radial pressure transforms under a disformal transformation as
\begin{align}
    \widetilde P_r = m_4\rho+m_5P_r+m_6P_\perp,
\end{align}
where the explicit expressions for $m_4$, $m_5$, and $m_6$
can be read off from the equations in Appendix~\ref{appB-dtemt}.
It is obvious that if $\rho(r_s)=P_r(r_s)=P_\perp(r_s)=0$ then
$\widetilde P_r$ vanishes there. Therefore, the notion of
the surface of a star is invariant under a disformal transformation.
By utilizing the consistency of the hydrostatic equation~\eqref{eq:hydrostatic},
we have thus arrived at the stronger conclusion than
that in Ref.~\cite{Minamitsuji:2021rtw}.
Note in passing that
in the aforementioned exceptional case where $\psi'(r_s)=0$,
we have $m_4=m_6=0$ and hence the same conclusion holds.

\section{The case of k-essence}\label{sec:caseofk}

Let us study the case in which the gravitational part of the Lagrangian is given by
\begin{align}
    {\cal L}_{\rm grav}=\frac{\mpl^2}{2}R+G_2(X),
\end{align}
i.e., the case of shift-symmetric k-essence disformally coupled to matter.
We require that the Minkowski spacetime is a solution of the theory.
The function $G_2$ then must be such that
\begin{align}
    G_2(X_*)=0,\quad G_{2X}(X_*)=0,
\end{align}
for some $X_*$ (see Eqs.~\eqref{eq:tr-k}--\eqref{eq:rr-k} below).\footnote{The
Minkowski specetime with $X=X_*$ is pathological because the sound speed of scalar
fluctuations vanishes~\cite{deRham:2019gha}. This issue can be remedied by introducing a
small higher-derivative term~\cite{Mukohyama:2005rw,Motohashi:2019ymr}.
This kind of small corrections are ignored in the solutions presented in this paper.}
The value of $X_*$ is thus determined from the parameters of the model
under consideration.

If $\Gamma=0$, the speed of gravitational waves in this theory coincides with
that of light. Moreover, the theory is free from graviton decay into dark
energy~\cite{Creminelli:2018xsv,Creminelli:2019nok} and dark energy instability induced by gravitational
waves~\cite{Creminelli:2019kjy}. Therefore, k-essence with $\Gamma=0$ is consistent with
gravitational wave observations~\cite{LIGOScientific:2017zic}.
In the following discussion, however, we will not restrict ourselves to
the case of $\Gamma=0$ for the sake of generality.

The $(tr)$-component of the gravitational field equations reads
\begin{align}
    {\cal G}_{tr}=-\mu \psi'G_{2X}=-\chi,
\end{align}
which, combined with Eq.~\eqref{eq:chih01}, gives
\begin{align}
    \mu \psi'(h-G_{2X})=0.\label{eq:tr-k}
\end{align}
Thus, we have two branches: $\psi'=0$ and $h-G_{2X}=0$.
We call the former the Branch I and the latter the Branch II.
The $(tt)$- and $(rr)$-components are given, respectively, by
\begin{align}
    {\cal G}_t^t&=-\frac{\mpl^2}{r^2}\left(
    1-e^{-\lambda}+re^{-\lambda}\lambda'
    \right)+2XG_{2X}-G_2=-\rho,\label{eq:tt-k}
    \\
    {\cal G}_r^r&=-\frac{\mpl^2}{r^2}\left(
    1-e^{-\lambda}-re^{-\lambda}\nu'
    \right)-G_2=P_r.\label{eq:rr-k}
\end{align}
Note that in the Branch I there is no effect of
the coupling functions $\Omega$ and $\Gamma$ on the profile of the metric functions.

\subsection{Interior solution}

Let us begin with the inspection of the interior solution.
We assume that the radial and tangential pressures are
related to $\rho$ through equations of state: $P_r=P_r(\rho)$ and $P_\perp=P_\perp(\rho)$.
To determine the metric functions in the stellar interior,
we first solve Eq.~\eqref{eq:tr-k} algebraically to write
$\psi'=F(\nu,\lambda,\rho)$.
(Obviously, $F=0$ in the Branch I.)
Using this, one can express $T_\phi$ in terms of
$\nu,\lambda,\rho$, and their first derivatives.
Thus eliminating $\psi'$,
Eqs.~\eqref{eq:hydrostatic},~\eqref{eq:tt-k}, and~\eqref{eq:rr-k}
reduce to three first-order differential equations for $\nu$, $\lambda$, and $\rho$.
Given the boundary conditions at $r=0$, one can integrate these equations
outwards to determine the stellar structure.
The radius of a star, $r_s$, is defined by $P_r(r_s)=0$.

We require that $\nu$, $\lambda$, $\psi$, $\rho$, $P_r$, and $P_\perp$
are regular at $r=0$. Then,
near the center, we may expanded these quantities as
\begin{align}
    \nu&=\nu_c+\frac{\nu_2}{2}r^2+\cdots,\label{expand1}\\
    \lambda&=0+\frac{\lambda_2}{2}r^2+\cdots,\\
    \psi'&=\psi_2 r+\dots,\label{expand3}\\
    \rho&=\rho_c+\frac{\rho_2}{2}r^2+\cdots,\\
    P_r&=P_{r,c}+\frac{P_{r,2}}{2}r^2+\cdots,\\
    P_\perp&=P_{\perp,c}+\frac{P_{\perp,2}}{2}r^2+\cdots.
\end{align}
Note here that, as can be seen from Eqs.~\eqref{eq:tt-k} and~\eqref{eq:rr-k},
we have $1-e^{-\lambda}={\cal O}(r^2)$ near the center and hence $\lambda(0)=0$.
In contrast, $\nu_c$ is not determined locally and 
one can integrate the equations for any value of $\nu_c$.
The choice of $\nu_c$ is correlated with how one sets the zero of
the gravitational potential. Usually, one tunes $\nu_c$ so that
the boundary conditions in the stellar exterior are satisfied.

From Eqs.~\eqref{expand1}--\eqref{expand3} we see that
\begin{align}
    X=X_c-\frac{1}{2}(X_c\nu_2+\psi_2^2)r^2+\cdots ,
\end{align}
where
\begin{align}
     X_c:=\frac{1}{2}\mu^2e^{-\nu_c}.   
\end{align}
Therefore, for a function $f$ of $X$ (such as $w_1$), we have
$f(X)=f(X_c)+{\cal O}(r^2)$ near the center.
Now, it is easy to see that $W^r={\cal O}(r)$ and $T_\phi=\;$constant
near the center. This fact and Eq.~\eqref{eq:hydrostatic} imply that
\begin{align}
    P_r-P_\perp={\cal O}(r^2)\quad \Rightarrow \quad 
    P_{r,c}=P_{\perp,c}\,(=:P_c).
\end{align}

On the basis of the above argument, let us turn back to Eq.~\eqref{eq:tr-k}.
It is clear that the Branch I
is always consistent with the boundary conditions at $r=0$.
In particular, in this branch,
$\mu$ is a free parameter as far as the interior solution is concerned.
In contrast, for a given central density $\rho_c$ (and a central pressure
according to the equation of state),
\begin{align}
    \left.h-G_{2X}\right|_{r=0}&=
    \frac{ w_1 P_c-w_2 (\rho_c-3P_c) +2X_cw_3 \rho_c}{1-2X_c w_1}
    \notag \\ & \quad 
    -G_{2X}(X_c)
\end{align}
is vanishing only if a particular value of $X_c$ is chosen.
This implies that the Branch II is consistent
only for a particular value of $\mu$, and this particular value is
dependent on the central density $\rho_c$ of a star.

\subsection{Exterior solution}

First let us consider the Branch I, in which
the interior solution with $\psi'=0$ is smoothly connected to
the exterior solution with $\psi'=0$.
In this branch one cannot obtain an analytic solution valid
in the entire region of the stellar exterior.
We therefore discuss linearized solutions around the
Minkowski spacetime with $X=X_*$:
\begin{align}
\nu &\simeq \nu_*+\delta \nu(r) \ll 1, \quad \lambda \simeq \delta \lambda(r) \ll 1,
\notag \\ 
X &\simeq \frac{\mu^{2}}{2}e^{-\nu_*}(1-\delta \nu),
\end{align}
where
\begin{align}
	\nu_*=\ln\left(\frac{\mu^2}{2X_*}\right).
\end{align}
By taking $\mu=\sqrt{2X_*}$, one can set $\nu_*=0$.
With this choice, $\mu$ is determined solely by theory parameters.

Linearization is justified sufficiently away from the star, $r\gg r_s$.
The linearized field equations read
\begin{align}
\frac{r \delta \lambda'+\delta \lambda}{r^{2}}+q \delta \nu=0, \quad r \delta \nu'-\delta \lambda=0,
\end{align}
where we defined
\begin{align}
    q:=\frac{2 X_{*}^{2} G_{2 X X}\left(X_{*}\right)}{M_{\mathrm{Pl}}^{2}}.
\end{align}
The general solution to these equations is given by
\begin{align}
\delta \nu&=C_{1} \frac{\cos (\sqrt{q} r)}{r}+C_{2} \frac{\sin (\sqrt{q} r)}{r},
\\
\delta\lambda&=-\sqrt{q}\left[C_1\sin(\sqrt{q}r)-C_2\cos(\sqrt{q}r)\right]-\delta \nu,
\end{align}
for $q>0$ and
\begin{align}
\delta \nu&=C_{3} \frac{e^{\sqrt{-q} r}}{r}+C_{4} \frac{e^{-\sqrt{-q} r}}{r},
\\
\delta\lambda&=\sqrt{-q}\left[C_3
e^{\sqrt{-q} r}-C_{4} e^{-\sqrt{-q} r}
\right]-\delta \nu,
\end{align}
for $q<0$.

The oscillatory behavior for $q>0$ is qualitatively similar to
what has been found in
the Newtonian potential in the ghost condensate model~\cite{ArkaniHamed:2003uy}.
Suppose that the length scale associated to $q$ is very large,
say, the cosmological horizon scale, and consider the metric
in the region $r_s\ll r\ll 1/\sqrt{q}$. We then have
\begin{align}
    \delta\nu\simeq C_{2}\sqrt{q}+\frac{C_{1}}{r},
    \quad 
    \delta\lambda\simeq -\frac{C_{1}}{r}.
\end{align}
After fixing $\mu$ as $\mu=\sqrt{2X_*}$, one can tune $\nu_c$
so that the boundary condition $C_2=0$ is satisfied.
The metric functions for $r_s\ll r\ll 1/\sqrt{q}$ are thus
described by the linearized Schwarzschild solution
with $-C_1$ being the mass parameter.

In the case of $q<0$, the diverging term can be removed by
imposing the boundary condition $C_3=0$.
After fixing $\mu$ as $\mu=\sqrt{2X_*}$, this can be satisfied
by tuning $\nu_c$. Then,
in the region $r_s\ll r\ll 1/\sqrt{-q}$, we have
\begin{align}
    \delta\nu\simeq -C_4\sqrt{-q}+\frac{C_4}{r},
    \quad \delta\lambda\simeq -\frac{C_4}{r}.
\end{align}
In this case one can no longer remove the constant part $-C_4\sqrt{-q}$,
but this term is physically unimportant.

Next, let us consider the Branch II, in which
the interior solution with $h-G_{2X}=0$ is connected to
the exterior solution with $G_{2X}=0$ at the surface where $h=0$.
In this branch,
we have $X=X_*$ in the entire region of the stellar exterior,
and the metric functions take the
Schwarzschild form,
\begin{align}
    e^\nu=e^{\nu_\infty}\left(1-\frac{C_0}{r}\right),\quad
    e^{-\lambda}=1-\frac{C_0}{r},\label{eq:stSch}
\end{align}
where $\nu_\infty$ and $C_0$ are integration constants.
Note that $\nu_\infty$ has no physical significance and can be set to zero
by shifting $\nu_c$.

\subsection{Numerical examples}

\begin{figure}[tb]
    \begin{center}
        \includegraphics[keepaspectratio=true,height=60mm]{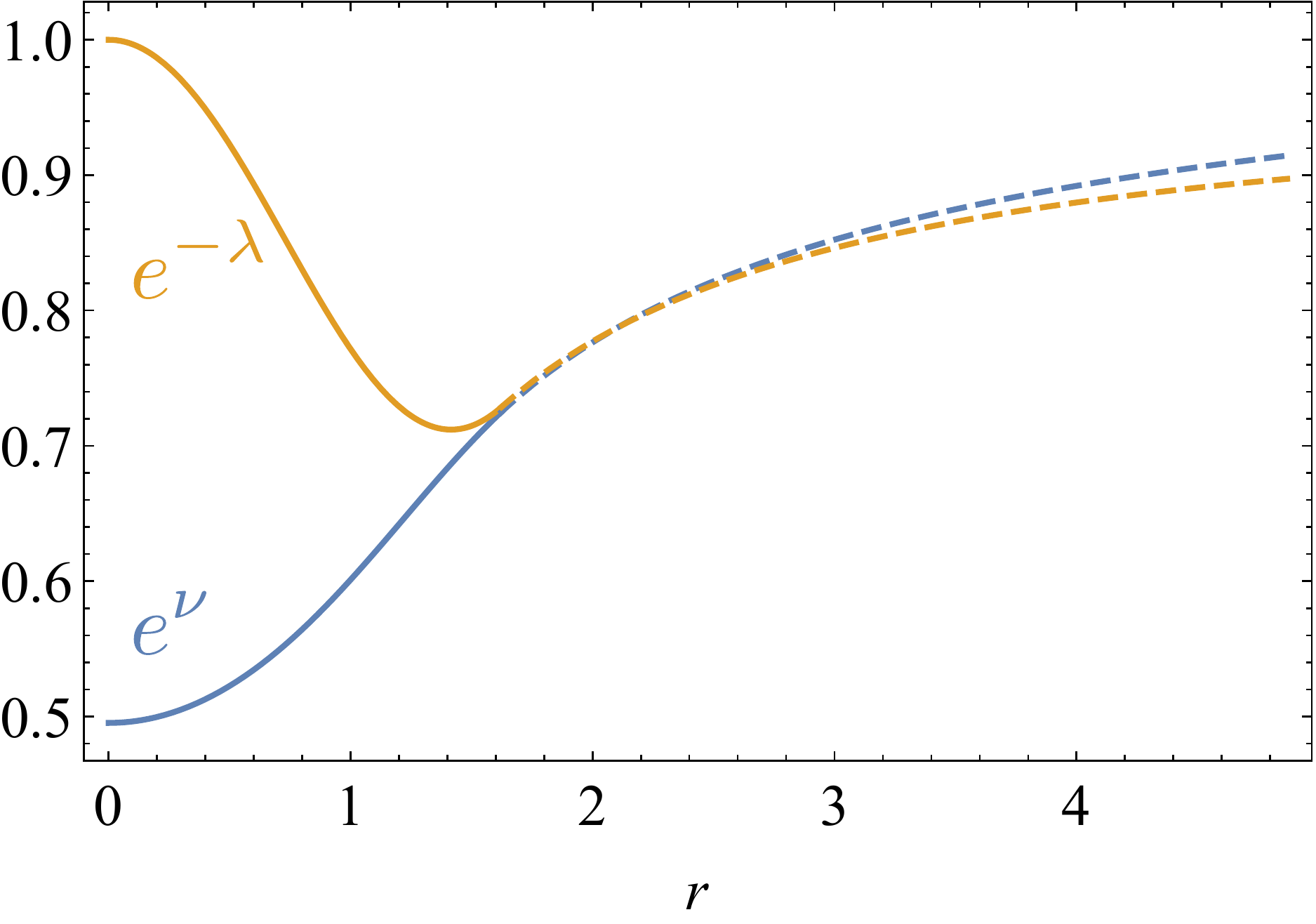}
    \end{center}
      \caption{
      The metric functions in the case where $q>0$ and $\mu=\sqrt{2X_*}$.
      The interior (exterior) region corresponds to solid (dashed) lines.
      The $r\lesssim 1/\sqrt{q}$ region is zoomed.
  	}
      \label{fig:fig_tr_01.pdf}
  \end{figure}

\begin{figure}[tb]
    \begin{center}
        \includegraphics[keepaspectratio=true,height=60mm]{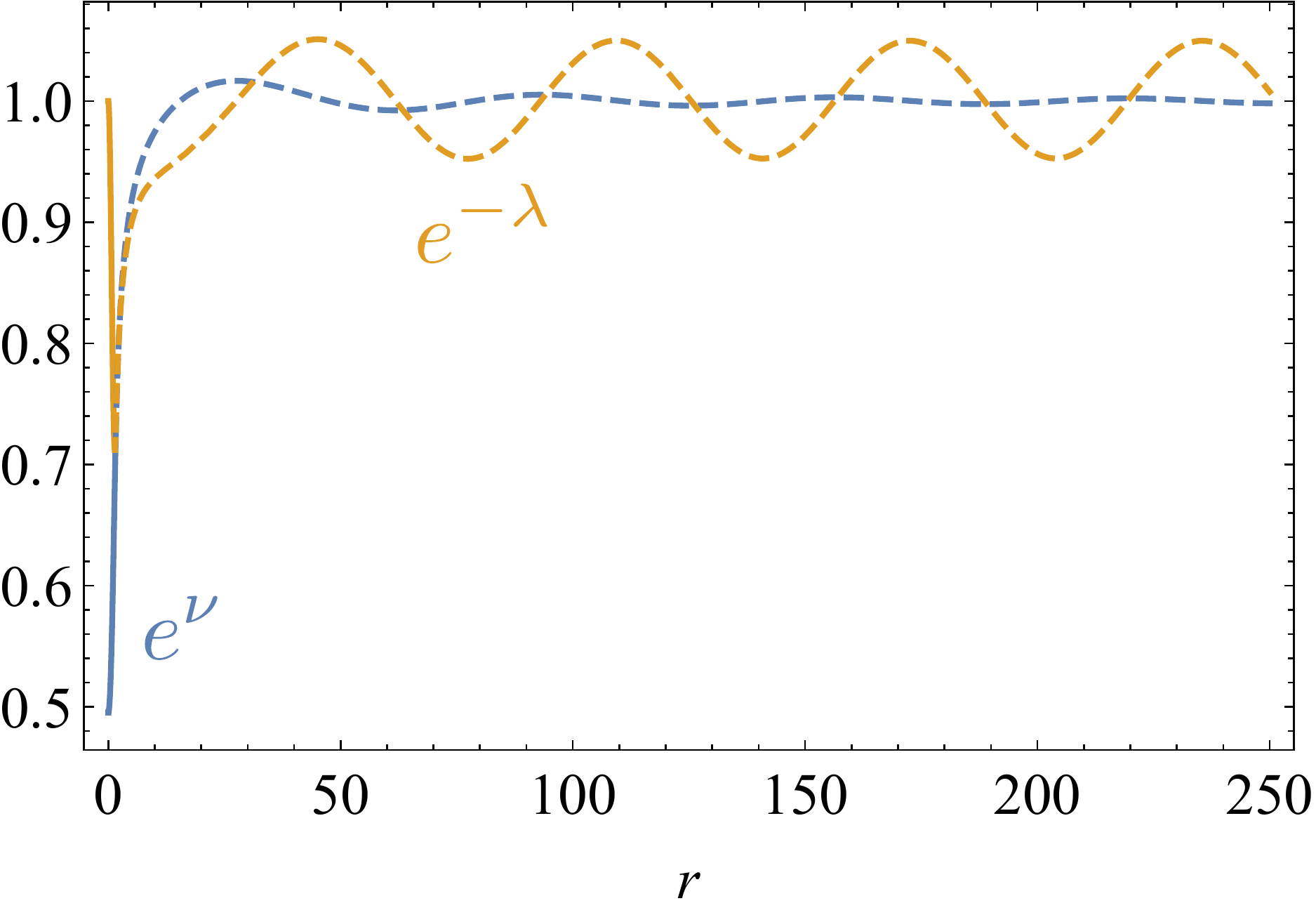}
    \end{center}
      \caption{
      The same as Fig.~\ref{fig:fig_tr_01.pdf}, but the plotted range is $r\gg 1/\sqrt{q}$
      so that the oscillatory behavior can be seen.
      The numerical solution is well approximated by
      $\nu\simeq -0.49 \cos(\sqrt{q}r)/r$ for $r\gtrsim 1/\sqrt{q}$.
  	}
      \label{fig:fig_tr_02.pdf}
  \end{figure}

\begin{figure}[tb]
    \begin{center}
        \includegraphics[keepaspectratio=true,height=60mm]{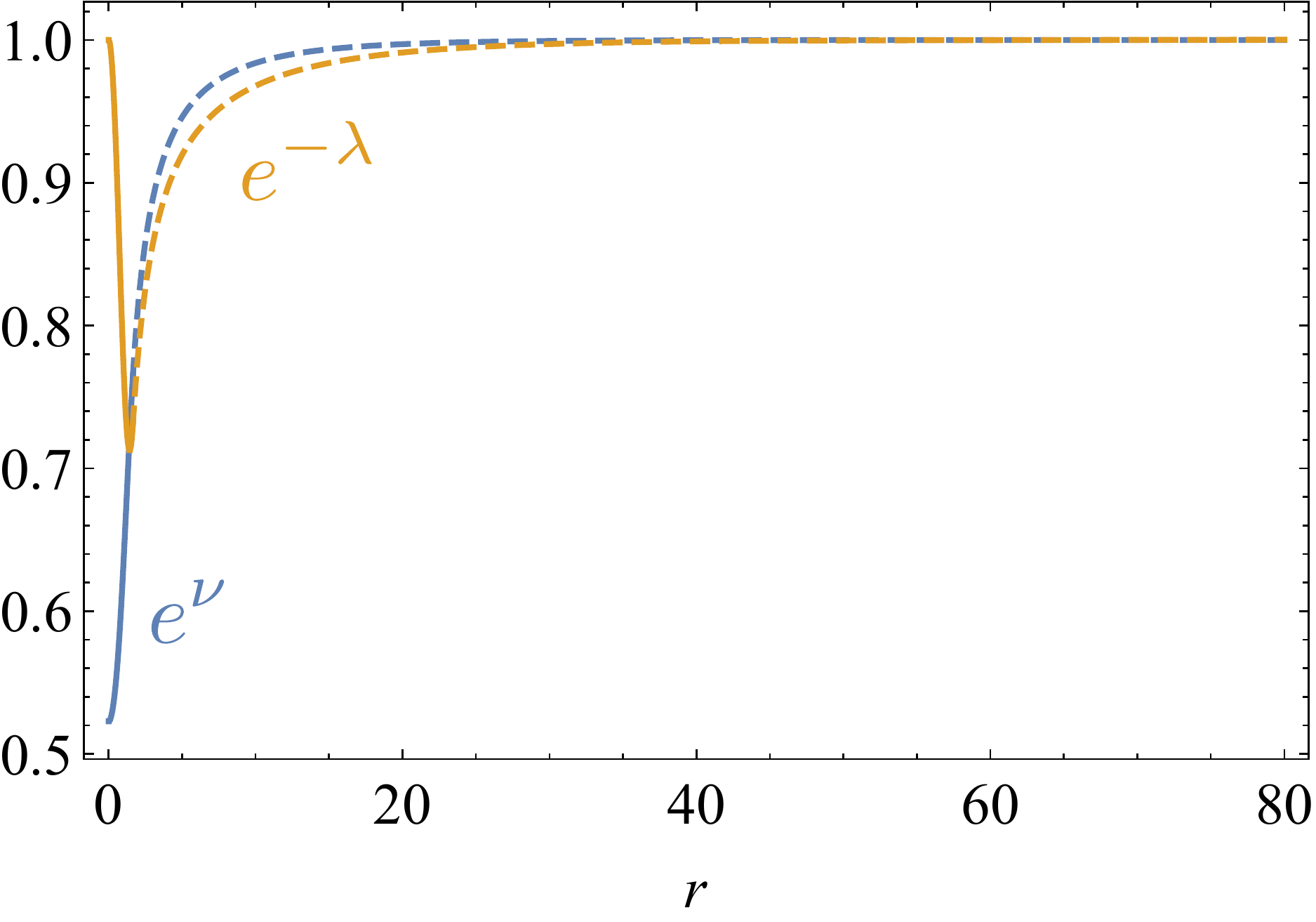}
    \end{center}
      \caption{
      The metric functions in the case where $q<0$ and $\mu=\sqrt{2X_*}$.
      The interior (exterior) region corresponds to solid (dashed) lines.
      The numerical solution is well approximated by
      $\nu\simeq -0.43 e^{-\sqrt{-q}r}/r$ for $r\gtrsim 1/\sqrt{-q}$.
  	}
      \label{fig:fig_tr_03.pdf}
  \end{figure}

\begin{figure}[tb]
    \begin{center}
        \includegraphics[keepaspectratio=true,height=60mm]{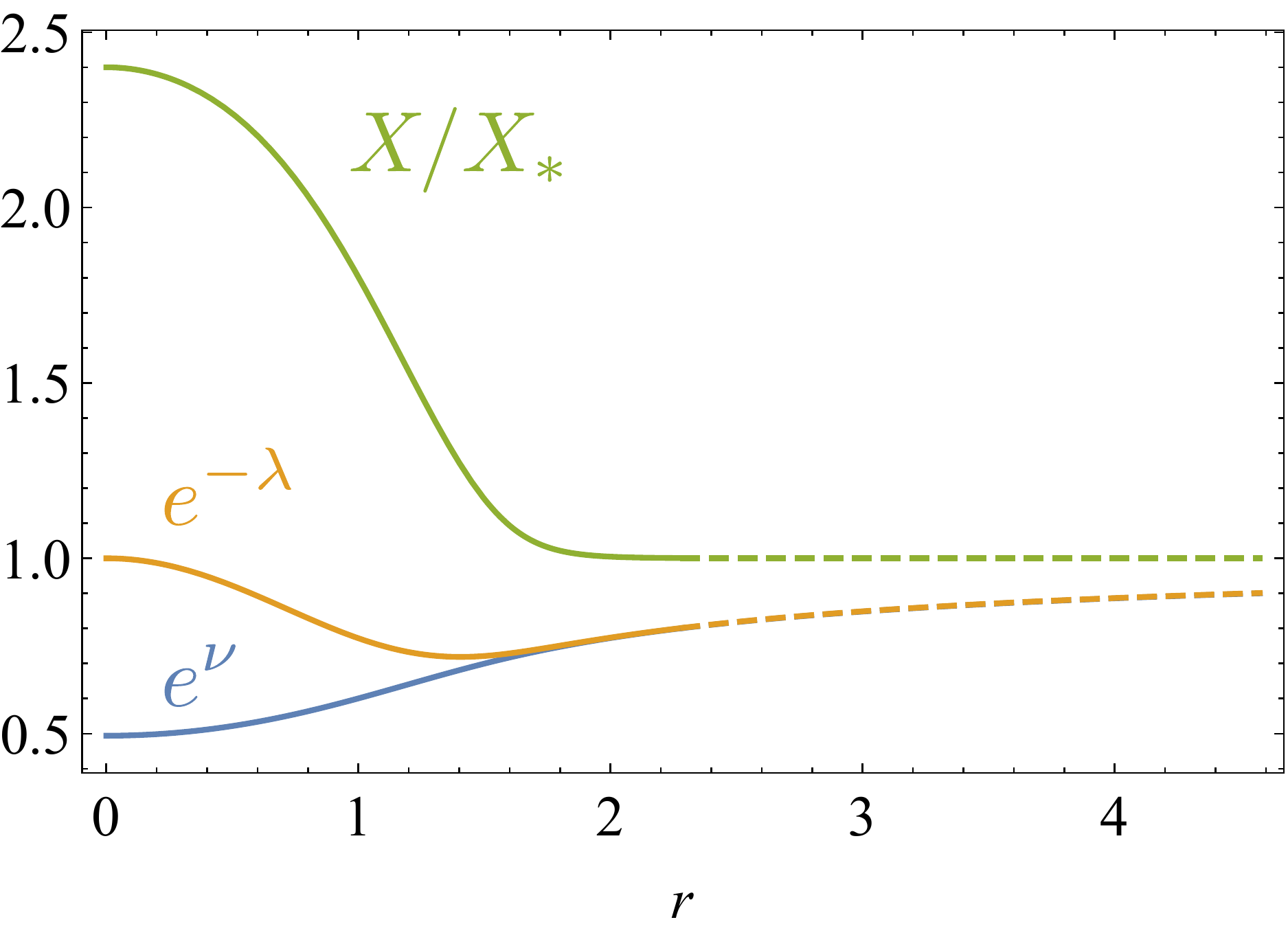}
    \end{center}
      \caption{
      The metric functions and $X/X_*$ in the case where
      $\mu$ is chosen so that $h-G_{2X}=0$ at the center.
      The interior (exterior) region corresponds to solid (dashed) lines.
  	}
      \label{fig:fig_nontr.pdf}
  \end{figure}

As a concrete example we consider a specific model with
\begin{align}
    G_2=\frac{q\mpl^2}{4}\frac{(X-X_*)^2}{X_*^2},
    \quad 
    \Omega=e^{2\beta X/X_*},\quad \Gamma=0,
\end{align}
and present our numerical results in Figs.~\ref{fig:fig_tr_01.pdf}--\ref{fig:fig_nontr.pdf}.
We assume for simplicity that the equation of state is isotropic and is given by
$P_r=P_\perp\propto \rho^2$,
where the constant of proportionality is fixed so that $P_c=0.1\rho_c$.
We use units where $\rho_c/\mpl^2=1$. The parameters are given by
$q=0.01$ for the solutions in Figs~\ref{fig:fig_tr_01.pdf} and~\ref{fig:fig_tr_02.pdf},
$q=-0.01$ in Fig.~\ref{fig:fig_tr_03.pdf}, and
$q=\beta=0.01$ in Fig.~\ref{fig:fig_nontr.pdf}.


The Branch I solution with $q>0$
is shown in Figs.~\ref{fig:fig_tr_01.pdf} and~\ref{fig:fig_tr_02.pdf},
where the behavior described in the previous subsection can be seen.
Similarly, the Branch I solution with $q<0$
is presented in Fig.~\ref{fig:fig_tr_03.pdf}.
Figure~\ref{fig:fig_nontr.pdf} shows the numerical example
of the Branch II. The interior solution is smoothly matched
to the exterior stealth Schwarzschild solution.

\subsection{Gravity that matter feels}

We have thus obtained the profile of the metric functions $\nu$ and $\lambda$.
However, since matter is coupled to the disformally-related
metric $\widetilde g_{\mu\nu}$,
the actual metric that we observe is $\widetilde g_{\mu\nu}$.
Therefore, let us write the line element
$\widetilde{\D s}{}^2=\widetilde{g}_{\mu\nu}\D \widetilde x^\mu\widetilde x^\nu$
in the form
\begin{align}
    \widetilde{\D s}{}^2&=
    -(1+2\widetilde\Phi)\D\widetilde \tau^2
    \notag \\ & \quad 
    +(1-2\widetilde{\Psi})
    [\D\widetilde\varrho^2+\widetilde\varrho^2(\D\theta^2+\sin^2\theta\D\varphi^2)],
    \label{dis-newton}
\end{align}
and evaluate $\widetilde\Phi$ and $\widetilde\Psi$ in the region
where linearization is valid.

In the Branch I, we have $X\simeq X_*(1-\delta\nu)$ and hence
$\Omega(X)=\Omega_*-X_*\Omega_{X*}\delta\nu$,
where the quantities with the asterisk are evaluated at $X=X_*$.
By performing the coordinate transformation
\begin{align}
    \widetilde \tau&=(\Omega_*-2X_*\Gamma_*)^{1/2}t,
    \\
    \widetilde \varrho&=\Omega^{1/2}_*\left(1+\frac{1}{2}
    \int^r\frac{\delta\lambda}{r'}\D r'\right)r,
\end{align}
we can recast the disformally-related metric to the form of
Eq.~\eqref{dis-newton} and find that
\begin{align}
    \widetilde\Phi&=\frac{\Omega_*-X_*(\Omega_{X*}-2X_*\Gamma_{X*})}{2(\Omega_*-2X_*\Gamma_*)}\delta\nu,
    \\
    \widetilde\Psi&=\frac{1}{2}\int^r\frac{\delta\lambda}{r'}\D r'
    +\frac{X_*\Omega_{X*}}{2\Omega_*}\delta\nu
    \notag \\ &=
    \frac{1}{2}\left(1+\frac{X_*\Omega_{X*}}{\Omega_*}\right)\delta\nu,
\end{align}
where we used $\delta\lambda=r\delta\nu'$ in the last line.
We therefore have
\begin{align}
    \frac{\widetilde\Psi}{\widetilde\Phi}
    =\frac{(1+X_*\Omega_{X*}/\Omega_*)(1-2X_*\Gamma_*/\Omega_*)}%
    {1-X_*(\Omega_{X*}-2X_*\Gamma_{X*})/\Omega_*}.\label{phi-psi-ratio}
\end{align}

This result can be used to evaluate the deviation from
general relativity. Note that it is easy to find the cases
where $\Omega_X$ and $\Gamma$ are not small, but $\widetilde\Psi\simeq \widetilde \Phi$.
A simple example is given by
\begin{align}
    \Omega=\Omega_0e^{\beta_1X/X_*},
    \quad 
    \Gamma = \frac{\beta_2\Omega_0}{X_*}e^{\beta_1X/X_*}.
\end{align}
In this case we see that $\widetilde\Psi\simeq \widetilde \Phi$
for $\beta_2\simeq \beta_1/(1+2\beta_1)$.
However, in the case of conformal coupling, $\Gamma=0$,
$X_*\Omega_{X*}/\Omega$ must be sufficiently small.

In the Branch II, the disformally-related metric
cannot take the form of Eq.~\eqref{dis-newton} in general.
Instead one needs to consider
the metric with a solid deficit angle~\cite{BenAchour:2019fdf},
\begin{align}
    \widetilde{\D s}{}^2&=
    -(1+2\widetilde\Phi)\D\widetilde \tau^2
    \notag \\ & \quad 
    +(1-2\widetilde{\Psi})
    [\D\widetilde\varrho^2+
    s_0^{-1}\widetilde\varrho^2(\D\theta^2+\sin^2\theta\D\varphi^2)],
\end{align}
where $s_0$ is a constant.
To put the metric into this form, we perform the coordinate transformation
\begin{align}
    \D\widetilde\tau &=(\Omega_*-\mu^2\Gamma_*)\left(\D t
    -\frac{\mu\Gamma\psi'}{\Omega e^\nu-\mu^2\Gamma}\D r\right),
    \\
    \widetilde\varrho&=
    s_0^{1/2}\Omega_*^{1/2}
    \left[1+
    \frac{1}{2}\int^r\left(\frac{\delta\lambda}{r'}
    -\frac{\mu^2\Gamma_*}{\Omega_*-\mu^2\Gamma_*}\frac{\delta\nu}{r'}\right)\D r'
    \right]r,
\end{align}
where
\begin{align}
    s_0:=\frac{\Omega_*-2X_*\Gamma_*}{\Omega_*-\mu^2\Gamma_*}.
\end{align}
It is then found that
\begin{align}
    \widetilde \Phi&=\frac{1}{2}\frac{\Omega_*}{\Omega_*-\mu^2\Gamma_*}\delta \nu,
    \\
    \widetilde\Psi &=
    \frac{1}{2}\int^r\left(\frac{\delta\lambda}{r'}
    -\frac{\mu^2\Gamma_*}{\Omega_*-\mu^2\Gamma_*}\frac{\delta\nu}{r'}\right)\D r'.
\end{align}
Since $\delta\nu=-\delta\lambda =-C_0/r$,
we have
\begin{align}
    \widetilde\Phi=\widetilde\Psi \propto \frac{1}{\widetilde\varrho}.
\end{align}
Thus, in the Branch II
the difference from general relativity
can be found in the solid deficit angle~\cite{BenAchour:2019fdf}.
Since the deficit angle changes geodesics, 
we can measure its effect by observing the light propagation,
e.g., the deflection angle of the light.~\cite{Barriola:1989hx,Ono:2018jrv}.

\section{Conclusions}\label{sec:concl}

In this paper, we have studied static and
spherically symmetric stars disformally coupled with a scalar field.
We have been focusing on shift-symmetric theories,
so that the conformal and disformal factors of the metric
that couples to matter are dependent only on
$X:=-g^{\mu\nu}\partial_\mu\phi\partial_\nu\phi/2$.
The shift symmetry admits the scalar-field configuration
that is dependent linearly on time,
$\phi=\mu t+\psi(r)$, where $\mu\,(=\partial_t\phi)$ is the constant
velocity of the scalar field.
Starting from a general scalar-tensor theory and
a general form of the energy-momentum tensor,
we have overviewed the gravitational field equations and
fluid equations for matter. More specifically, our energy-momentum
tensor allows for the radial heat flux and anisotropic pressure.
Such a general form of the energy-momentum tensor
is implied by the way it transforms under a disformal transformation,
and hence is natural to consider in the present setup.

We have investigated the density and pressure profile
in the vicinity of the surface of a star (defined as the location
at which the radial pressure vanishes, $P_r(r_s)=0$).
It is known that the density $\rho$ and tangential pressure $P_\perp$ may be
discontinuous across the stellar surface as long as
matter is minimally coupled to gravity.
In the presence of $X$-dependent conformal and disformal factors,
however, we have shown that $\rho(r_s)=P_\perp(r_s)=0$
is requried by the structure of the hydrostatic equilibrium equation.
This in turn indicates the disformal invariance of
the surface of a star. This fact was previously subtle and
shown only when certain extra conditions are met~\cite{Minamitsuji:2021rtw}.

We have then concentrated on the shift-symmetric k-essence theory
disformally coupled to matter, and
studied the interior and exterior metric functions and scalar-field profile.
We have found that there are two branches of the solution
depending on the value of $\mu$.
One branch, which we call the Branch I,
is characterized by the trivial scalar-field profile, $\psi'=0$.
The velocity $\mu$ in the Branch I is determined solely by theory parameters.
The other branch, which we call the Branch II,
is characterized by the stealth Schwarzschild geometry with
the constant $X$ profile in the exterior region.
The velocity $\mu$ in the Branch II is determined by the central density
as well as theory parameters.
The disformally transformed metric that couples to matter
has also been discussed.
We have found in particular
that in the Branch I the standard result of general relativity
can be reproduced if a certain relation between the conformal and
disformal factors is satisfied.

As we have seen in Fig.~\ref{fig:fig_tr_02.pdf},
the Branch I solution exhibits an oscillatory behavior
of the metric functions far away from the source,
which may lead to interesting cosmological consequences.
We hope to come back to the study of cosmology in the present setup in the near future.

\acknowledgments
We are grateful to Daisuke Yamauchi for fruitful discussions.
TI especially thanks to Tomohiro Harada and Yu Nakayama for
their critical comments on his master's thesis defense presentation.
The work of TI was supported by the Rikkyo University Special Fund for Research.
The work of AI was supported by the JSPS Research Fellowships for Young Scientists No.~20J11285.
The work of TK was supported by
MEXT KAKENHI Grant Nos.~JP20H04745 and~JP20K03936.

\appendix

\section{Derivation of Eqs.~\eqref{eq:Tphi=WZ}--\eqref{eq:defz2}}\label{app:basic-eqs}

We start with writing $T^{\mu\nu}$ as
\begin{align}
    T^{\mu\nu}
    &=\frac{\partial \widetilde g_{\alpha\beta}}{\partial g_{\mu\nu}} 
    \cdot 
    \frac{2}{\sqrt{-g}}\frac{\partial}{\partial \widetilde g_{\alpha\beta}}
    \left(\sqrt{-\widetilde g}{\cal L}_{\rm m}\right)
    \notag \\ 
    &={\cal J}_g\biggl(
    \Omega \widetilde T^{\mu\nu}+\frac{1}{2}\Omega_X\phi^\mu\phi^\nu 
    g_{\alpha\beta}\widetilde T^{\alpha\beta} 
    \notag \\ &\quad 
    +\frac{1}{2}\Gamma_X\phi^\mu\phi^\nu \phi_\alpha\phi_\beta\widetilde T^{\alpha\beta} 
    \biggr),\label{app:eq:tilt}
\end{align}
where
\begin{align}
    {\cal J}_g&=\frac{\sqrt{-\widetilde g}}{\sqrt{-g}}=\Omega^{3/2}
    \sqrt{\Omega-2X\Gamma},
    \\
    \widetilde T^{\mu\nu}&=\frac{2}{\sqrt{-\widetilde g}}
    \frac{\partial}{\partial \widetilde g_{\mu\nu}}\left(
    \sqrt{-\widetilde g}{\cal L}_{\rm m}
    \right).\label{def:T-tilde}
\end{align}
By contracting Eq.~\eqref{app:eq:tilt} with $g_{\mu\nu}$ and $\phi_\mu\phi_\nu$
and solving the resultant equations for $g_{\mu\nu}\widetilde T^{\mu\nu}$
and $\phi_\mu\phi_\nu \widetilde T^{\mu\nu}$,
we obtain 
\begin{align}
    g_{\mu\nu}\widetilde T^{\mu\nu} &=
    \frac{(\Omega+2X^2\Gamma_X)T+X\Gamma_X\phi_\mu\phi_\nu T^{\mu\nu}}%
    {{\cal J}_g\Omega (\Omega-X\Omega_X+2X^2\Gamma_X)},
    \\
    \phi_\mu\phi_\nu \widetilde T^{\mu\nu} &=
    \frac{-2X^2\Omega_XT+(\Omega-X\Omega_X)\phi_\mu\phi_\nu T^{\mu\nu}}%
    {{\cal J}_g\Omega (\Omega-X\Omega_X+2X^2\Gamma_X)}.
\end{align}
Thus, we have
\begin{align}
    {\cal J}_g\Omega\widetilde T^{\mu\nu}&=T^{\mu\nu} 
    -\frac{\Omega_X\phi^\mu\phi^\nu T}{2(\Omega-X\Omega_X+2X^2\Gamma_X)}
    \notag \\ &\quad 
    -\frac{\Gamma_X\phi^\mu\phi^\nu T^{\alpha\beta}\phi_\alpha\phi_\beta}%
    {2(\Omega-X\Omega_X+2X^2\Gamma_X)}.\label{app:eq:tilt-t}
\end{align}

Now, we may write
\begin{align}
    T_\phi = -\frac{1}{\sqrt{-g}}\frac{\delta S_\phi}{\delta\phi}
    =\nabla_\mu W^\mu-Z,
\end{align}
where 
\begin{align}
    W^\mu&=\frac{1}{\sqrt{-g}}\frac{\partial}{\partial\phi_\mu} 
    \left(\sqrt{-\widetilde g}{\cal L}_{\rm m}\right)
    =\frac{1}{2}\frac{\partial \widetilde g_{\alpha\beta}}{\partial \phi_\mu} 
    \cdot{\cal J}_g\widetilde T^{\alpha\beta},\label{app:eq:defW}
    \\
    Z&=\frac{1}{\sqrt{-g}}\frac{\partial}{\partial\phi} 
    \left(\sqrt{-\widetilde g}{\cal L}_{\rm m}\right)
    =\frac{1}{2}\frac{\partial \widetilde g_{\alpha\beta}}{\partial \phi} 
    \cdot{\cal J}_g\widetilde T^{\alpha\beta}.\label{app:eq:defZ}
\end{align}
Substituting Eq.~\eqref{app:eq:tilt-t} to Eqs.~\eqref{app:eq:defW}
and~\eqref{app:eq:defZ}, we arrive at Eqs.~\eqref{eq:wmTTT}--\eqref{eq:defz2}.

\section{Disformal transformation of the energy-momentum tensor}\label{appB-dtemt}

Here we summarize the disformal transformation of each component of
the energy-momentum tensor that composes a star. One can find
a similar discussion for a slowly rotating star
dressed with a time-{\em independent} scalar field
in Ref.~\cite{Minamitsuji:2021rtw}.

For a static and spherically symmetric metric,
Eq.~\eqref{intro:disf} reads
\begin{align}
    \widetilde g_{\mu\nu}\D x^\mu\D x^\nu 
    &=-\left(\Omega e^\nu-\mu^2\Gamma\right)\D t^2+2\mu \Gamma\psi'\D t\D r
    \notag \\ &\quad
    +\left[\Omega e^\lambda+\Gamma(\psi')^2\right]\D r^2
    \notag \\ &\quad 
    +\Omega r^2
    \left(\D\theta^2+\sin^2\theta\D\varphi^2\right).
\end{align}
Introducing the new coordinates defined by
\begin{align}
    \D \widetilde t&=
    \D t-\frac{\mu\Gamma\psi'}{\Omega e^\nu-\mu^2\Gamma}\D r,
    \\ 
    \widetilde r&=\Omega^{1/2} r,
\end{align}
the metric can then be put into the standard diagonal form as
\begin{align}
    \widetilde g_{\mu\nu}\D \widetilde x^\mu\D \widetilde x^\nu 
    &=-e^{\widetilde \nu}\D\widetilde t^2 
    +e^{\widetilde\lambda}\D\widetilde r^2+
    \widetilde r^2
    \left(\D\theta^2+\sin^2\theta\D\varphi^2\right),
\end{align}
where
\begin{align}
    e^{\widetilde\nu}&=\Omega e^\nu-\mu^2\Gamma,
    \\
    e^{\widetilde\lambda}&=\frac{\Omega-2X\Gamma}{\Omega-\mu^2e^{-\nu}\Gamma}
    \left(1+\frac{rX'\Omega_X}{2\Omega}\right)^{-2}e^{\lambda}.
\end{align}

Note that we do not always have
$\widetilde\nu\to 0$ and $\widetilde\lambda\to 0$
as $\widetilde r\to\infty$ even if
$\nu\to0$ and $\lambda\to 0$ as $r\to\infty$.
One can rescale the coordinates so that 
$\widetilde\nu\to 0$ and $\widetilde\lambda\to 0$
as $\widetilde r\to\infty$ in terms of the rescaled coordinates,
but rescaling the radial coordinate
then results in a solid deficit angle~\cite{BenAchour:2019fdf}.

Now one can see that
the energy-momentum tensor $\widetilde T_\mu^\nu$ defined by
Eq.~\eqref{def:T-tilde} is diagonal:
\begin{align}
    \widetilde T_{\widetilde\mu}^{\widetilde\nu} =
    {\rm diag}\left(
    -\widetilde\rho,\widetilde P_r,\widetilde P_\perp,\widetilde P_\perp
    \right),
\end{align}
i.e., $\widetilde T^{\widetilde t\widetilde r}$ vanishes.
To show this we used Eq.~\eqref{eq:chih01}.
Using Eq.~\eqref{app:eq:tilt-t} we obtain
\begin{align}
    \widetilde\rho &=\frac{M_1\rho+M_2P_r+M_3P_\perp}{{\cal J}_g%
    \{1-\mu^2e^{-\nu}[w_1 +2w_3e^{-\lambda}(\psi')^2]\}},
    \\
    \widetilde P_r&=\frac{M_4\rho+M_5P_r+M_6P_\perp}{{\cal J}_g%
    \{1-\mu^2e^{-\nu}[w_1 +2w_3e^{-\lambda}(\psi')^2]\}}, 
    \\
    \widetilde P_\perp&=
    {\cal J}_g^{-1}P_\perp,
\end{align}
where 
\begin{align}
    M_1&=\frac{1}{\Omega-X\Omega_X+2X^2\Gamma_X}\times 
    \notag \\ & \quad \times \biggl[
    \Omega\left(1-e^{-\nu}\mu^2\frac{\Gamma}{\Omega}\right)^2+\frac{1}{2}
    e^{-2\lambda}(\psi')^4\Gamma_X
    \notag \\ & 
    +\frac{e^{-\lambda}(\psi')^2}{2}
    \left(1-2e^{-\nu}\mu^2\frac{\Gamma}{\Omega}+2e^{-\nu}\mu^2X\frac{\Gamma^2}{\Omega^2}
    \right)\Omega_X
    \biggr],
    \\
    M_2&=-\frac{e^{-\nu}\mu^2}{2(\Omega-X\Omega_X+2X^2\Gamma_X)}\times 
    \notag \\ & \quad \times \biggl[
    e^{-\lambda}(\psi')^2\left(2\frac{\Gamma^2}{\Omega}+\Gamma_X\right)
    \notag \\ & \qquad 
    +\left(1-2e^{-\nu}\mu^2\frac{\Gamma}{\Omega}+2e^{-\nu}\mu^2X\frac{\Gamma^2}{\Omega^2}
    \right)\Omega_X
    \biggr],
    \\
    M_3&=2e^{-\nu}\mu^2w_2
    \frac{\Omega-2X\Gamma-2e^{-\lambda}(\psi')^2\Gamma}{\Omega},
    \\
    M_4&=e^{-\lambda}(\psi')^2(-w_2+e^{-\nu}\mu^2w_3),
    \\
    M_5&=\frac{\Omega-2X\Gamma}{\Omega-X\Omega_X+2X^2\Gamma_X}
    +
    e^{-\nu}\mu^2(w_2-e^{-\nu}\mu^2w_3),
    \\
    M_6&=2e^{-\lambda}(\psi')^2w_2.
\end{align}
This result implies that even if the fluid is isotropic in one frame,
it is anisotropic in the other frame, as already
pointed out in Ref.~\cite{Minamitsuji:2021rtw}.
By taking $\mu=0$ one can reproduce the expressions
presented in Ref.~\cite{Minamitsuji:2021rtw}.

\section{Jump between branches}

In Sec.~\ref{sec:caseofk} we have found two branches of the solution
on the basis of the $(tr)$-component of the gravitational field equations~\eqref{eq:tr-k}.
In this appendix,
we show that a transition from one branch to another can not
occur in the interior region of the star unless
an exceptional fine tuning takes place.
The discussion here is similar to that of Ref.~\cite{Lehebel:2017fag}.

Suppose that a solution exhibits a transition from
the Branch I ($\psi'=0$) to the Branch II ($h-G_{2X}=0$) at some $r=r_0$.
(The transition from the Branch II to the Branch I can be
treated in the same way.)
We assume that $\nu, \lambda, \psi, \rho, P_r$, and $P_\perp$ are $C^1$.
We can then expand various quantities around $r_0$:
\begin{align}
&\nu \simeq \nu_{0}+\nu_{1}\left(r-r_{0}\right), \quad 
\lambda \simeq\lambda_{0}+\lambda_{1}\left(r-r_{0}\right), \notag \\
&\rho \simeq \rho_{0}+\rho_{1}\left(r-r_{0}\right), \quad 
P_r \simeq P_{r,0}+P_{r,1}\left(r-r_{0}\right),\quad 
\notag \\ &
P_\perp \simeq P_{\perp,0}+P_{\perp,1}\left(r-r_{0}\right).
\end{align}
Since our field equations include functions of $X$,
it should be expanded similarly as
\begin{align}
    X\simeq  X_0+X_1(r-r_0),
\end{align}
so that for any smooth function $f$ of $X$ we have $f\simeq f(X_0)+f'(X_0)X_1(r-r_0)$.
This implies that
\begin{align}
    &\psi'=0\quad \qquad \qquad \qquad (r\le r_0),
    \\
    &\psi'\simeq \psi_{1/2}'(r-r_0)^{1/2}\quad (r\ge r_0).
\end{align}
Now it can be seen that Eq.~\eqref{eq:hydrostatic} contains
\begin{align}
    T_\phi\psi' &= \frac{e^{-(\nu+\lambda) / 2}}{r^{2}}\left(r^{2} 
    e^{(\nu-\lambda) / 2} \psi' h\right)'\psi'
    \notag \\ &=
    \left(\psi'\right)^{2} \frac{e^{-(\nu+\lambda) / 2}}{r^{2}}\left(r^{2} e^{(\nu-\lambda) / 2} h\right)'+\psi' \psi'' e^{\lambda} h,
\end{align}
and the second term in the last line
shows a discontinuity at $r=r_0$. However,
this is not acceptable because 
all the other terms in the same equation are $C^1$.

One possible loophole is the case where $\psi_{1/2}'=0$ and
the expansion of $\psi'$ starts with a higher-order term in $r-r_0$.
Another loophole is the case where $h=0\,(=G_{2X})$ at $r=r_0$.
In both cases, an extreme fine tuning is required to
the theory, the equation of state, and the central density of the star.
Only under such an extreme fine tuning, a jump between
the branches could occur, and hence practically we do not need
to care about the possibility of a branch jump.

\bibliography{refs}
\bibliographystyle{JHEP}
\end{document}